\documentclass{epl}

\title{
ELECTRON--PHONON COUPLING AND ANHARMONIC EFFECTS
IN METAL CLUSTERS}
\author{F.F. Karpeshin\inst{1,2}
\and J. da Provid\^encia\inst{1}
\and C. Provid\^encia\inst{1}
\and J. da Provid\^encia jr.\inst{3}}

\institute{ \inst{1}Departamento de F\'\i sica, Universidade de
Coimbra -  3004-516 Coimbra, Portugal\\
  \inst{2} Institute of Physics, St.Petersburg University -
198904 St.Petersburg, Russia \\
\inst{3} Departamento de F\'\i sica, Universidade da Beira
Interior -  6201-001 Covilh\~a, Portugal} \pacs{36.40.-c}{
Atomic and molecular clusters} \pacs{36.40.Gk}{
Plasma and collective effects in clusters} \pacs{36.40.Qv}{
Stability and fragmentation of clusters}\pacs{36.40.Wa}{
Charged clusters}

\begin{document}

\maketitle

\begin{abstract}
  The periods of the harmonic oscillations of the ion core of charged sodium
clusters around the equilibrium shapes are considered. It is found that these
periods
are of the order of magnitude of the experimentally measured relaxation times
of the plasmons, which suggests the importance of
the electron-ion coupling
and stresses the role played by the electron-phonon
interaction
in the dissipation of the plasmon energy.
The relation of the process to fission is briefly discussed.

\end{abstract}
\hyphenation{ioni-zation}
\newcommand{\beq}{\begin{equation}}
\newcommand{\eeq}{\end{equation}}
\newcommand{\beqa}{\begin{eqnarray}}
\newcommand{\eeqa}{\end{eqnarray}}
\newcommand{\oh}{\frac{1}{2}}
\newcommand{\gapprox}{_{\displaystyle>}\atop
^{\displaystyle\sim}}
\newcommand{\lapprox}{_{\displaystyle<}\atop
^{\displaystyle\sim}}
\newcommand{\widespacing}{\baselineskip 5ex}
\newcommand{\shortspacing}{\baselineskip 3ex}
\def\d{{\rm d}}
\def\a{\alpha}

\section{ Introduction}

A considerable progress in cluster physics has been achieved for
the past years. The picture of the collective motion of the
electron system is well understood, \cite{joao,moseler,rh}. The
collective electron modes have also been studied experimentally.
On the other hand, experiment shows a strong electron-ion
correlation, with a characteristic time of $\sim 10^{-12}$ s
\cite{helm}. There is no complete picture to explain such a strong
correlation occurring in spite of the huge difference in the
masses which actually justifies the use of the Born-Oppenheimer
approximation. In principle, the energy of a plasmon can be
directly transferred to the phonons  within a time which follows
from the analytic solution of a simple model to be presented in
greater detail elsewhere, \beq\tau_{pl}=\pi\left({ M\over
m}\right)^{1\over2}\omega_{pl}^{-1},\label{tau} \eeq where
$\omega_{pl}$ is the plasmon frequency, {$M$ is the mass of the
ion  and $m$ is the electron mass}. For sodium clusters, this time
is of the order of 10$^{-13}$s, which may be even an order of
magnitude shorter than the experiments show. The relaxation time
of the plasmon may be identified with $\tau_{pl}$, eq.(\ref{tau}).
{ We summarize now the derivation of eq. (\ref{tau}). We consider
plane waves propagating along the z axis. We denote the
displacements of the electronic and ionic distributions,
respectively, by $u=u(z,t)$ and $v=v(z,t)$. We denote the
fluctuations of the
electronic 
and 
ionic charge densities, respectively, by $\delta\rho_e=e{\partial
u\over \partial z}$ and
 $\delta\rho_i=-e{\partial v\over \partial z}$.
The Lagrangian describing the plasmon--lattice dynamics reads
\beqa &L=&\int\d z\left({1\over 2}M\left({\partial v\over \partial
t}\right)^2+ {1\over 2}m\left({\partial u\over \partial
t}\right)^2-
{1\over 2}\a\left({\partial v\over \partial z}\right)^2\right)\nonumber\\
&&-4\pi  e^2\int\d z\int\d z'\left({\partial u\over\partial z}-{\partial v\over\partial z}\right)|z-z'|
\left({\partial u'\over\partial z'}-{\partial v'\over\partial z'}\right)\nonumber
\eeqa
where $u'=u(z'), \,v'=v(z'),$ $m$ is the electron mass, $M$ is the ion mass and
 ${1\over 2}\a({\partial v\over \partial z})^2$ is the elastic energy density of
the lattice.
The equations of motion read
\beqa &&-M\frac{\partial^2v}{\partial
t^2}+\a\frac{\partial^2v}{\partial z^2} +4\pi e^2({ u}-{
v})=0\nonumber
\\
&& -m\frac{\partial^2u}{\partial t^2}+4\pi
e^2({v}-{u})=0\,. \eeqa
It follows that, for an appropriate value of the wave vector,
the number of plasmons at time $t$ is
\beq N_{pl}=uu^*\propto\cos^2\left({1\over2}\sqrt{m\over M}\omega_{pl}t\right),
\eeq
so that  the plasmon energy will be completely transferred to the
lattice in time $\tau_{pl}.$ }

Our present purpose is to
draw attention to the fact that the periods of the
collective vibrations of the ionic core remarkably coincide with
the detected electron relaxation times. This fact supports the
assumption that the interaction of the electron system with the
collective ionic modes plays {a rather} essential part in
dissipation, as, e.g., in the electron-phonon interaction in
crystals. There is no need to 
assume that the phonon excitation will necessarily
result into fission, though the two aspects are generically
related to one another, so that fission is the ultimate form of a
superposition of many of phonons 
in the limit of
large amplitude oscillation 
\cite{I}.
 The importance of the collective
modes for the electron relaxation was also noted in ref.
\cite{gerchik}. The influence of the ionic degrees of
freedom on the electronic excitations has been considered in \cite{moseler,rh,bertch}.

In turn, cluster fission is a process of great interest.
Experiment shows predominance of the strongly asymmetric fission
accompanied by emission of mono-, di- or, more rarely, trimers.
Symmetric fission remains among the most important topics of
research. Study of symmetric fission allows
one to better understand 
the dynamics of the interplay
between single-particle and collective degrees of freedom
\cite{I}.

        In paper \cite{II},
we have considered the interplay between rotational and
vibrational modes of the collective motion in clusters, finally
leading to fission. It was specifically found that rotation favors
fission through phase transitions occurring in the shape of the
clusters rotating with large
angular momentum. This is similar to 
nuclear fission. On the other hand, only
neutral clusters were studied in \cite{II}. It is well
known, however, that the charge of the clusters 
plays an  important role 
in fission.

In the present paper we 
extend our considerations to 
charged clusters. Moreover, we compare the obtained results  with
the experimental data, which became available after paper
\cite{II} was submitted for publication. Sodium
 clusters with $N=18$, 43, 92 and
470  atoms  are considered.
In the light of recent experiments
\cite{helm}, we discuss the
role played by the collective motion of the core 
in the electron-ion correlations.
{Our discussion is based on the LDM (see \cite{LDM} and references
cited in \cite{I}) which is enough for the present purposes. We
take into account surface and Coulomb energies. The inclusion of a
shell-correction term generally allows one to obtain a detailed
description of specific features of clusters, such as binding
energy per particle, ionization energy etc. \cite{I}. Here, we
leave out these aspects, focusing rather on general tendencies
than on detailed descriptions of particular properties of
individual clusters. On the other hand, the LDM is
well-suited for the description of fission, due to its 
intuitive appeal and transparency. Many properties of nuclear
fission have been understood in the framework of the LDM.
Representative sodium clusters in a large range of the number $N$
of the constitutive atoms is dealt with, one of which, $N=92$, is
close to the clusters studied experimentally in \cite{helm}.
We also draw special attention to possible cases of soft clusters
with charge close to the critical value, just
at the border of Coulomb  stability, namely, clusters with
$N=43,\, q=3$ and $N=470,\, q=10,$ $q$ being the cluster charge.

\section{Outline of the model}
{ We consider
oscillations  of an incompressible liquid drop
such that, at each instant,  the shape is that of  
an axially-symmetric spheroid, with time-dependent half-axis $c$
in the direction $z$, and time-dependent half-axes $a=b$ in the
perpendicular directions $x$ and $y$.  We assume irrotational
flow. In view of volume conservation, the values of $c$ and $b$
are related, at each instant, by $cb^2 = R^3$, $R$ being the
radius of the equilibrium spherical shape. For sodium,  $R = 3.93
N^{1/3}$bohr, $N$ being the number of the atoms in the cluster
\cite{ArVi}. We choose $c$ as the collective variable. The mass
parameter can be found by solving the Laplace equation for the
velocity field with appropriate boundary condition, analogously to
the case of small-amplitude multi-pole vibration \cite{bohr}: \beq
\Delta \chi (x,y,z) = 0 \;\;\;\;.\label{1} \eeq The velocity field
is \beq {{\mathbf v}}(x,y,z) =  - \nabla \chi (x,y,z)
\;\;\;.\label{2} \eeq A solution of eq. (\ref{1}) satisfying the
proper boundary condition is \beq\chi=-{\dot c \over 4
c}(x^2+y^2-2z^2)\;.\label{6} \eeq The vibrational kinetic energy
can be found as follows: \beq T = \oh \mu \int (\nabla \chi )^2 \d
V ,\label{8} \eeq where $\mu$ denotes the mass per unit volume.
From Eq. (\ref{6}) one gets the expression
\beq
T = \oh {\cal M} (c) \dot{c}^2\,,\quad  \label{9}
{\cal M}(c) = \frac{1}{5} M (1 + \oh u^3) \;\;\;,
\eeq
where $M$ is the total mass of the cluster, and $u$
stands for $(R / c)$.}

The potential energy of deformation arises from the interplay of
the opposite effects of the restoring surface and the repulsive
Coulomb energies. For the {surface} energy, the following
expression was
derived in \cite{II}, 
 \beqa
 V_{Surf}(c) &=& 2 \pi \sigma R^2 \left[u^{-1/2} {{\arcsin
\sqrt{1 -
u^3} } \over {\sqrt{1 - u^3}}} + u\right] - 4 \pi
\sigma R^2
\;\;\;\;\;\;\;\;
\mbox{for} \;\;\;\;R<c
\;\;\;,   \nonumber \\
\label{11}\\
V(c)_{Surf} &=& 2 \pi \sigma R^2  \left[u^{-1/2} {{\mbox{arcsh}
\sqrt{u^3 - 1} }
\over {\sqrt{ u^3 - 1}}} + u\right] - 4 \pi \sigma
R^2 \;\;\;\;\;\;\;\;
\mbox{for} \;\;\;\;R > c\;\;\;, \nonumber \eeqa where  $\sigma$
is the surface tension. For sodium clusters, {
we take $\sigma=3.8\times 10^{-3}$
 eV / bohr$^2$
in agreement with} the Stabilized Jellium Model \cite{LDM1,ArVi}.
The capacities of a prolate spheroid with half axes $c>a=b$ and of
an oblate spheroid with half axes $a=b>c$ are \cite{landau},
respectively, \beq C={\sqrt{c^2-a^2}\over \cosh^{-1}(c/a)}, \quad
{\rm and}\quad C={\sqrt{a^2-c^2}\over \cos^{-1}(c/a)}. \eeq These
formulae allow us to obtain the Coulomb energy of a charged
spheroid, \beq V_{Coul}(c)={q^2\over C}. \eeq For small
deformations the potential energy becomes \beqa
&V^{(2)}(c) &= V_{surf}^{(2)} + V_{Coul}^{(2)}        \\
&V_{surf}^{(2)}(c) &=
 \oh k_{surf} (c - R)^2 \\
&V_{Coul}^{(2)}(c) &= -\oh k_{Coul} (c - R)^2 \,, \eeqa where terms of
higher order than the second in $(c-R)$ have been neglected and
\beq
k_{surf} = \frac{16\pi\sigma}{5}=0.0382eV / a_0^2\,,    \quad
k_{Coul} = \frac{q^2 }{Na_0^3 r_s^3} = 0.1794eV
\frac{q^2}{Na_0^2}\,. \eeq Here, $a_0$ is the Bohr radius, $r_s =
3.93$, the mean radius per particle being $r_sa_0,$ and $q = 0, 1,
2,\cdots$ is the charge of the cluster.
As in \cite{II}, the kinetic energy
of the oscillating cluster is
\beq T = \frac{1}{2}{\cal M}(0)
{\dot{c}}^2\,, 
\eeq
 where ${\cal M}(0) = 6.417 \cdot 10^9eV N\, $ and
the oscillation period is
\beq
  \tau = 2 \pi \sqrt{\frac{\cal M}{k_{surf}-k_{Coul}}}\,.
\eeq For small amplitudes, our assumptions result in the
quadrupole vibrations of a liquid drop. If these assumptions are
used in conjunction with different velocity fields from the one
considered in the present paper, other types of fission will be
favored, as, for instance, asymmetric fission.

{\footnotesize
\begin{table}[!ht]
\caption{\small{ Behavior of sodium clusters with 18, 43, 92 and
470 atoms. In the second column, the charge $q$ of the cluster is
shown. For each cluster type, the constants $k_{surf}$ and
$k_{Coul}$, characterizing, respectively, the surface and the
Coulomb energies, and the period $\tau$ of small amplitude
oscillations around stable configurations, are presented.} }
\begin{center}
\begin{tabular}{||c|c||c|c|c|c||} \hline \hline
 Species  & q     &
  $k_{surf}\,(eVa_0^{-2}) $&   $k_{Coul}\,(eVa_0^{-2})$& $k_{surf} - k_{Coul}$ & $\tau$\,( s)  \\
\hline  \hline
  Na$_{18}$ &  0  &  0.382E-01  & 0         &  0.382E-01 &   0.193E-11    \\
        &     1   &  0.382E-01  & 0.997E-02  &  0.282E-01 &   0.224E-11    \\
\hline
       &      0   &  0.382E-01  & 0          &  0.382E-01 &   0.298E-11    \\
 Na$_{43}$ &  1   &  0.382E-01  & 0.417E-02  &  0.340E-01 &   0.316E-11    \\
        &     2   &  0.382E-01  & 0.167E-01  &  0.215E-01 &   0.397E-11    \\
        &     3   &  0.382E-01  & 0.375E-01  &  0.651E-03 &   0.228E-10    \\
\hline
        &     0   &  0.382E-01  & 0         &  0.382E-01 &   0.436E-11    \\
        &     1   &  0.382E-01 &  0.195E-02 &   0.362E-01&    0.448E-11   \\
 Na$_{92}$ &  2   &  0.382E-01 &  0.780E-02 &   0.304E-01&    0.489E-11   \\
        &     3   &  0.382E-01 &  0.175E-01 &   0.206E-01&    0.593E-11   \\
        &     4   &  0.382E-01 &  0.312E-01 &   0.700E-02&    0.102E-10   \\
\hline
     &        0   &  0.382E-01 &  0         &   0.382E-01&    0.985E-11   \\
     &        3   &  0.382E-01 &  0.344E-02 &   0.348E-01&    0.103E-10   \\
     &        5   &  0.382E-01 &  0.954E-02 &   0.287E-01&    0.114E-10   \\
 Na$_{470}$ & 7    & 0.382E-01 &  0.187E-01 &   0.195E-01&    0.138E-10   \\
      &       8    & 0.382E-01 &  0.244E-01 &   0.138E-01&    0.164E-10   \\
       &      9    & 0.382E-01 &  0.309E-01 &   0.728E-02&    0.226E-10   \\
       &     10    & 0.382E-01 &  0.382E-01 &   0.298E-04&    0.353E-09   \\
\hline  \hline
\end{tabular}
\end{center}
\label{clt1}
\end{table}
}


\section{ Results and discussion}

    The results of the calculation are presented in Table \ref{clt1}
for typical cluster with numbers of atoms 
$N = 18,$ 43, 92 and 470, 
the charge $q$ being such that  the clusters stay %
within the stability limit.

The periods can be compared with the experimental data
\cite{helm}. In that paper, characteristic times for the
relaxation of the plasmon energy were 
obtained, which turned out to be $2.5\times 10^{-12}$ s. As one
can see from Table \ref{clt1}, the plasmon relaxation time is
about the same as the characteristic period of the collective
vibration. Naturally, for the same period, the plasmon makes
thousands of oscillations.
The plasmon collective energy is quite likely dissipated by
inducing the excitation of the cluster phonons.
{The electrons collective energy is directly transferred to the
ion core, exciting phonon degrees of freedom. In view of the
relation $\omega_{ph} \ll \omega_{pl},$ the number of excited
phonons is very large, $n \gg 1$ \cite{I}. Such many-phonon
coherent excitations can then serve as doorway states for fission
\cite{doorw}, unless they are destroyed by dissipation. The latter
process dominates if the spreading width $\Gamma_s$ is greater
than the mean spacing of the $n$-phonon states, $\hbar
\omega_{ph}$, \beq \Gamma_s {\gapprox} \hbar \omega_{ph}\;\;\;.
\label{mf10} \eeq In the case of nuclear fission, the opposite
relation to (\ref{mf10}), namely, $\Gamma_s {\lapprox} \hbar
\omega_{ph}$, takes place \cite{doorw}. In this case, the
$n$-phonon structure is experimentally observable. This
corresponds to the Frenkel picture of fission \cite{frenkel},
where the collective deformation arises from a large amplitude
oscillation, which, due to an-harmonic effects, never comes back
to the
initial position, but proceeds towards fission. In this picture, 
fission occurs on a competitive basis with dissipation of the
phonon collective energy inside the ion core. The observed fact
that fission most likely occurs through the very asymmetric mode
of evaporation of monomers or dimers means that strong dissipation
of the core collective motion takes place, and that the relation
(\ref{mf10}) is satisfied. 
Moreover, from the expected phonon dissipation time $\tau_{diss}
\sim 10^{-12}$s, we can estimate the related spreading width of
the $n$-phonon states 
which
turns out to be \beq
\Gamma_s^{cl} 
\sim 7\times 10^{-4} eV \;. \label{mf11} \eeq That is, the
many-phonon states strongly overlap. This means that classical
approaches, such as the present one, are appropriate for their
description.

    We note that in some cases, if the cluster is near the border of
stability, in so far as $k_{Coul}$ approaches $k_{surf}$, the
vibration period
increases by an order of magnitude,
as for instance in the cases of Na$_{43}^{3+}$ and Na$_{470}^{10+}$ in
Table \ref{clt1}. One can expect that, under these circumstances,
the collective mode can be
more easily excited. Experimental study of the electron relaxation
and search for near to symmetric fission for such clusters would
be of the highest interest.
}

 \acknowledgments

    The authors would like to express their gratitude to H. Krappe,
V.O. Nesterenko, P.G. Reinhardt, A.V. Solovyov for helpful remarks.
They are particularly grateful to Hellmut Haberland for valuable
discussions of the experimental data.

    This work was partially supported by  FCT and FEDER under the
projects POCTI/FIS/451 /94, POCTI/35308/FIS/2000,
and NATO science fellowship.

\end{document}